\documentstyle[sprocl]{article}

\input{psfig}
\newcommand{\beq}{\begin{equation}}
\newcommand{\eeq}{\end{equation}}
\newcommand{\beqa}{\begin{eqnarray}}
\newcommand{\eeqa}{\end{eqnarray}}
\newcommand{\ba}{\begin{array}}
\newcommand{\ea}{\end{array}}

\begin{document}

\title{LARGE SHELL MODEL CALCULATIONS FOR CALCIUM ISOTOPES: 
SPECTRAL STATISTICS AND CHAOS}

\author{V.R. MANFREDI}

\address{Dipartimento di Fisica "G. Galilei", Universit\`a di Padova,\\
Istututo Nazionale di Fisica Nucleare, Sezione di Padova,\\
Via Marzolo 8, I 35131 Padova, Italy\\E-mail: manfredi@padova.infn.it} 

\author{J.M.G. GOMEZ}

\address{Departamento de Fisica Nuclear, Universidad Complutense de Madrid,\\
Avenida Complutense, E 28040 Madrid, Spain\\E-mail: gomezk@nuc1.fis.ucm.es}

\author{L. SALASNICH}

\address{Istituto Nazionale per la Fisica della Materia, Unit\`a di Milano,\\
Dipartimento di Fisica, Universit\`a di Milano, \\
Via Celoria 16, I 20133 Milano, Italy\\E-mail: salasnich@hpmire.mi.infm.it}

\maketitle\abstracts{We perform large shell model calculations 
for Calcium isotopes in the full $fp$ shell by using 
the realistic Kuo--Brown interaction. 
The Calcium isotopes are especially interesting because the nearest--neighbour 
spacing distribution $P(s)$ of low--lying energy levels shows significant 
deviations from the predictions of the Gaussian Orthogonal Ensemble 
of random--matrix theory. This contrasts with other neighbouring nuclei 
which show fully chaotic spectral distributions. 
We study the chaotic behaviour as a function of 
the excitation energy. In addition, a clear signature of 
chaos suppression is obtained when the single--particle spacings are 
increased. In our opinion the relatively weak strength of the 
neutron-neutron interaction is unable to destroy the 
regular single--particle mean--field motion completely. 
In the neighbouring nuclei with both protons 
and neutrons in valence orbits, on the other hand, 
the stronger proton-neutron interaction would appear to be sufficient 
to destroy the regular mean--field motion.}

\section{Introduction}
\par
The nearest neighbour spacing distribution of energy levels provides 
a good signature of the chaoticity of a quantum system. In fact, 
the fluctuation properties of quantum systems 
with underlying classical chaotic behaviour and 
time--reversal symmetry agree with the predictions of the Gaussian 
Orthogonal Ensemble (GOE) of random matrix theory, whereas 
quantum analogues of classically integrable systems display the features 
of the Poisson statistics\cite{b1}. 
\par
The nuclear shell model, with a realistic interaction and large configuration 
space, is one of the best theoretical approaches to the study of nuclear  
spectra. The model provides large sets of exact 
energy levels and wave functions 
in truncated space, and their statistical analysis can give information 
on the features and borderlines of the transition from regular to 
chaotic dynamics in nuclei. 
The statistical analysis of shell--model energy 
spectra and wave functions has mainly concentrated on the $sd$ 
shell region and chaotic behaviour has been found for 
these nuclei both near the yrast line and at higher energies\cite{b2}. 
\par 
In this paper we extend our recent statistical analysis of the shell--model 
energy levels in the $fp$ shell\cite{b3}, in order to study whether an order 
to chaos transition can be observed, and how it depends on 
variables such as excitation energy, angular momentum and 
single--particle energy spacings. 

\section{Calculations in the $fp$ shell}
\par
The nuclear shell--model Hamiltonian, in second--quantization notation, 
can be written as 
\beq
H=\sum_{\alpha} \epsilon_{\alpha}a_{\alpha}^+a_{\alpha} 
+{1\over 4}\sum_{\alpha \beta \gamma \delta} <\alpha \beta|V|\delta \gamma >
a_{\alpha}^+ a_{\beta}^+ a_{\gamma} a_{\delta} \; ,
\eeq 
where the labels denote the accessible single--particle states, 
$\epsilon_{\alpha}$ is the corresponding single--particle energy, 
and $<\alpha \beta|V|\delta \gamma >$ is the two--body matrix element 
of the nuclear residual interaction. 
\par 
We follow the standard approach to obtain the eigenvalues of the 
shell--model Hamiltonian. Exact calculations for several nuclei 
are performed in the 
($f_{7/2}$, $p_{3/2}$, $f_{5/2}$, $p_{1/2}$) 
configuration space, assuming a $^{40}$Ca inert core. 
The construction and diagonalization of large shell--model matrices 
are performed using 
a modified version of the computer code ANTOINE\cite{b4}. 
For a fixed number of valence protons and neutrons 
we calculate the energy spectrum for 
total angular momentum $J$ and total isospin $T$. 
The interaction is a minimally modified Kuo--Brown 
realistic force with monopole improvements\cite{b5}. Coulomb effects are not
included. 
\par
To analyze the energy level fluctuations, it is necessary to consider
only levels which have the same symmetries. In our case this means the
same number of nucleons and the same total angular momentum, parity and 
isospin. The level spectrum is then mapped onto unfolded levels with
quasi--uniform level density. The suitable unfolding procedure depends on
the region of the spectrum to be analyzed. 
For the low--lying levels, we use an unfolding procedure based on
the constant temperature formula, where 
the mean level density can be assumed to be of the form 
\beq
{\bar \rho}(E)={1\over T}\exp{[(E-E_0)/T]} ,
\eeq
with $T$ and $E_0$ fitting parameters\cite{b6}. 
We have compared this unfolding method 
with the standard local unfolding method\cite{b7}.
Provided that the number of energy levels is not too small, 
the two procedures give similar results in the low--energy region, but
the constant temperature level density is  smoother and is preferable 
in the ground--state region. When the analysis includes many levels
or the full spectrum, we use the local unfolding 
because, as is well known, due to the finite size of the shell--model 
basis, the eigenvalues are generally Gaussian distributed.

\section{Analysis of spectral statistics} 

The $P(s)$ distribution of the nearest--neighbour spacings 
$s_i={\bar N}(E_{i+1})-{\bar N}(E_i)$ of the unfolded levels 
is the best spectral statistics to study the fluctuations 
of the short range correlations. 
To quantify the chaoticity of $P(s)$ in terms of a parameter, 
we compare it to the Brody distribution, 
\beq
P(s,\omega)=\alpha (\omega +1) s^{\omega} \exp{(-\alpha s^{\omega+1})} \; ,
\eeq
with 
\beq
\alpha = (\Gamma [{\omega +2\over \omega+1}])^{\omega +1} \; . 
\eeq
This distribution interpolates between the Poisson distribution ($\omega =0$) 
of regular systems and the 
Wigner distribution ($\omega =1$) of chaotic ones (GOE). 
The parameter $\omega$ can be used as a simple 
quantitative measure of the degree of chaoticity\cite{b8}. 

\begin{table}[t]
\caption{Brody parameter $\omega$ for the nearest 
neighbour level spacing distribution 
for $0\leq J\leq 9$, $T=T_z$ states up to $4$, $5$ and $6$ MeV above the 
yrast line in the analysed nuclei.}
\vspace{0.2cm}
\begin{center}
\footnotesize
\begin{tabular}{|ccc|} \hline\hline 
Energy & $^{46}$V+$^{46}$Ti+$^{46}$Sc & $^{46}$Ca+$^{48}$Ca+$^{50}$Ca \\ 
\hline
$\leq 4$ MeV & 0.92 & 0.56\\ 
$\leq 5$ MeV & 0.93 & 0.60\\ 
$\leq 6$ MeV & 0.95 & 0.61\\ 
\hline\hline
\end{tabular}
\end{center}
\end{table}

\par
We analyze the energy spectra looking at the lowest 
energy region, up to a few MeV above the yrast line. The main 
problem in this region is that the number of energy levels 
of the same symmetry is too small for a reliable statistical analysis. 
But having at our disposal the whole energy spectrum, we can use a 
sufficiently large set of levels to determine the secular 
behaviour of the level density and perform the unfolding 
procedure for each symmetry class of states. Moreover, in order to 
obtain more meaningful statistics, after the unfolding we can combine 
the level spacings of different $J$ in a nucleus, or even of different 
nuclei, to calculate the $P(s)$ distribution. 
\par
A good estimate of the Brody parameter is obtained combining  
spacings of different nuclei. Table 1 shows the results for 
$^{46}$V+$^{46}$Ti+$^{46}$Sc and $^{46}$Ca+$^{48}$Ca+$^{50}$Ca, 
up to $6$ MeV above the yrast lines. 
The number of level spacings is now sufficiently large to yield 
meaningful statistics and we see that Ca isotopes are not very 
chaotic at low energy, in contrast to other nuclei in the same region. 
\par
The combined data of $^{46}$Ca+$^{48}$Ca+$^{50}$Ca show that 
the chaoticity increases slightly with excitation energy 
up to $6$ MeV. It is generally believed that states at higher energy, 
in the high density region, should be much more chaotic in nature.   
Thus, calcium isotopes offer the possibility to study 
the energy dependence of the chaoticity in the framework of the 
shell model. 

\begin{figure}[t]
\psfig{figure=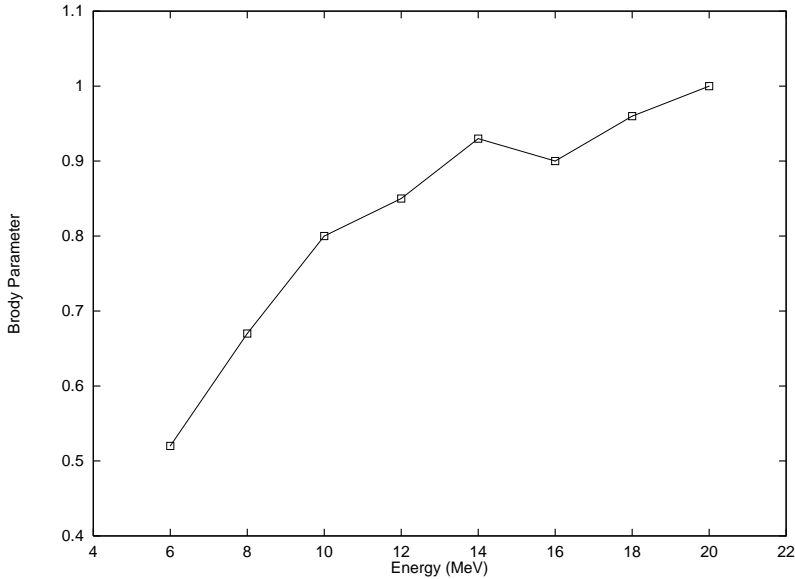,height=3.0in}
\caption{Energy dependence of the Brody parameter $\omega$ 
in $^{48}$Ca. All levels up to an energy $E$ 
above the yrast line are included for each $J$ value.}
\end{figure}

\par
We have calculated the $P(s)$ distribution and the Brody parameter 
up to a fixed value of the excitation energy above the yrast lines 
from $6$ to $20$ MeV. We use the local unfolding method 
and include all levels with $J=0$--$11$. We find that the chaoticity 
increases rather smoothly with energy. This is illustrated 
in Figure 1 for $^{48}$Ca. It is necessary to include levels 
up to about $14$ MeV to get $\omega > 0.9$. 
\par
Table 2 shows the Brody parameter $\omega$ for the whole spectrum 
of the analysed Ca isotopes, which range from $^{44}$Ca to $^{50}$Ca.  
We see that the lightest Ca isotopes are not fully chaotic even 
when the whole energy spectrum is taken into account. 
\par
For the heavier calcium isotopes the number of states is very large, 
e.g. $17,276$ for $^{50}$Ca. Therefore, it is possible to analyze 
separately the spectra for different $J$ values with good statistics.  
We do not find any significant dependence on angular momentum, 
except for $J=0$ where we find $\omega\simeq 0.8$ both for 
$^{48}$Ca and $^{50}$Ca. This is probably related to the 
fact that the pairing interaction, which preserves the seniority 
quantum number, is more effective for the $J=0$ states. 

\begin{figure}[t]
\psfig{figure=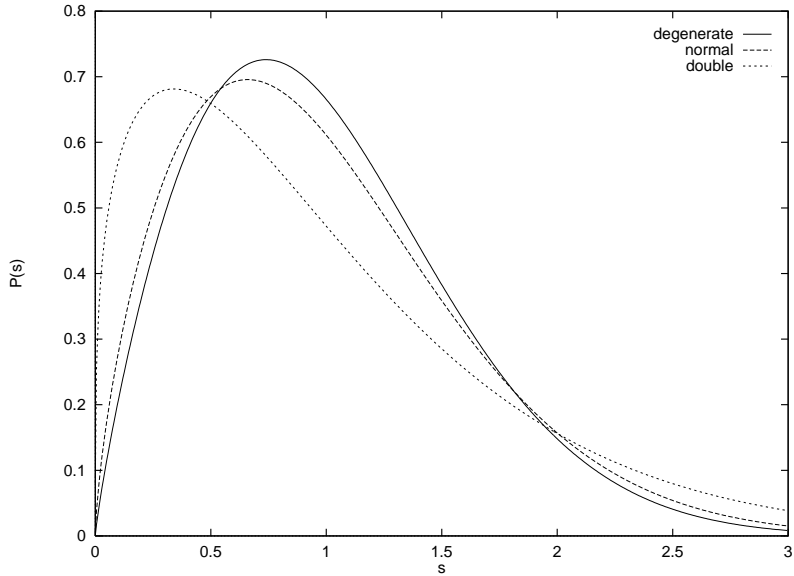,height=3.0in}
\caption{$P(s)$ distribution for $^{44}$Ca. 
All levels for each $J$ value are included. The single--particle 
energies are: degenerate, normal and double spaced.}
\end{figure}

\par 
It is very interesting to study the effect of the one--body Hamiltonian on 
the $P(s)$ distribution. The single--particle motion in the spherical mean
field is regular, while the nuclear two--body residual interaction is strongly 
non--linear. Figure 2 shows how the fluctuation properties of nuclear
energy levels in $^{44}$Ca change when the single--particle
spacings are changed. We consider three cases: i) the
$f_{7/2}$, $p_{3/2}$, $f_{5/2}$ and $p_{1/2}$ 
single--particle spacings are all degenerate; ii) the 
single--particle spacings are the experimental ones; 
iii) the single--particle spacings are multiplied by a factor of two. 
The realistic residual interaction is the same in all cases. 
The $P(s)$ distribution includes 
all the nuclear levels, separately unfolded for each $J$ value.
For degenerate single--particle levels Figure 2 clearly shows chaotic
dynamics in $^{44}$Ca, but when the single--particle spacings are increased, 
there is a transition towards regularity. 
The same kind of chaos to order transition is observed for the other 
Calcium isotopes. 

\begin{table}[t]
\caption{Brody parameter $\omega$ for the nearest neighbour 
level spacing distribution for Ca isotopes. 
All levels for each $J$ value are included.} 
\vspace{0.2cm}
\begin{center}
\footnotesize
\begin{tabular}{|ccccccc|} \hline\hline 
$^{44}$Ca & $^{45}$Ca & $^{46}$Ca & $^{47}$Ca & $^{48}$Ca 
& $^{49}$Ca & $^{50}$Ca \\ 
\hline
0.69 & 0.75 & 0.99 & 0.98 & 0.95 & 1.00 & 0.87 \\ 
\hline\hline
\end{tabular}
\end{center}
\end{table}

\section{Conclusions}
\par
The Ca isotopes are especially interesting because the nearest--neighbour 
spacing distribution $P(s)$ of low--lying energy levels shows significant 
deviations from GOE predictions, in contrast to other neighbouring nuclei 
which show fully chaotic spectral distributions.
\par
The analysis of level spacings up to a given excitation energy shows that 
the chaoticity of Ca isotopes increases smoothly with the 
excitation energy end point. However, 
even when the whole energy spectrum is included, 
the lighter isotopes $^{44}$Ca and $^{45}$Ca are not fully chaotic, and  
the order to chaos transition is progressive as the number of
active particles increases.
\par
In the heavier Ca isotopes studied, the number of energy levels is very 
large and it is possible to analyse with good statistics the energy levels 
of a single J value. For example, in $^{50}$Ca the dimension of many 
Hamiltonian matrices exceeds one or two thousand for fixed J. We do not 
find significant differences in the $P(s)$ distribution for different J 
values, except perhaps for $J=0$, for which we obtain a Brody parameter  
somewhat smaller than for other $J$ values. This is probably related to 
the pairing component of the interaction and the aproximate
conservation of the seniority quantum number.
\par
Finally, a clear chaos to order transition is observed 
as the single--particle spacings are 
increased. In the lighter Ca isotopes we obtain chaotic behaviour for 
degenerate single--particle states and quasi--regular motion when the 
single--particle spacings are twice the experimental values. Thus, it seems 
that the main reason for the substantial deviations from chaoticity 
obtained in Ca isotopes is that the relatively weak strength of the 
neutron-neutron interaction is not able to destroy the 
regular single--particle mean--field motion completely. But in the 
neighbouring nuclei with both protons and neutrons in valence orbits, the  
stronger proton-neutron interaction seems to be sufficent to destroy the
regular mean--field motion.

\end{document}